\documentclass[aps,prl,amsmath,twocolumn,superscriptaddress,letterpaper,floatfix]{revtex4-1}
\usepackage{graphicx,color}
\usepackage{verbatim}
\usepackage{amssymb}   
\usepackage{amsmath}
\usepackage{amsfonts}
\usepackage{mathdots}
\usepackage{hyperref}
\usepackage{epsfig}

\usepackage{braket}
\usepackage{bm}

\begin{document}

\title{Real-Space Interference from  Quantum Metric}
\author{Bo-Qing Wu}

\affiliation{Department of Physics, Hong Kong University of Science and Technology, Hong Kong, China}

\author{Xing-Lei Ma}
\email{xmabk@connect.ust.hk}

\affiliation{Department of Physics, Hong Kong University of Science and Technology, Hong Kong, China}

\author{Jin-Xin Hu}

\affiliation{Department of Physics, Hong Kong University of Science and Technology, Hong Kong, China}

\author{Jinchao Zhao}

\affiliation{Department of Physics, Hong Kong University of Science and Technology, Hong Kong, China}

\author{K. T. Law}
\email{phlaw@ust.hk}
\affiliation{Department of Physics, Hong Kong University of Science and Technology, Hong Kong, China}

\date{\today}

\begin{abstract}
Real-space interference of wave packets is governed by their central momentum separation and momentum width, which set the oscillation wavelength and spatial envelope, respectively. Here, we show that in crystalline systems, quantum metric distribution in an isolated flat band can similarly give rise to real-space
interference through the evanescent tails of localized bound states. 
The location and width of each quantum metric hot spot in momentum space identify an evanescent wave component of a bound state. With multiple hot spots, these components interfere in real space, producing oscillatory patterns in both the density distribution of a single bound state and the coupling strength between spatially overlapping bound states. The oscillation period is set by the momentum-space separation between the hot spots, while the long-range decay length is determined by the width of the narrowest hot spot. Furthermore, we demonstrate that the interference can be experimentally probed using Josephson scanning tunneling microscopy, where the Josephson critical current oscillates with the tip position. With momentum-resolved quantum geometry becoming experimentally accessible, our results provide a direct physical interpretation of its local momentum-space structure: quantum-metric hot spots encode the oscillation and decay scales governing long-distance real-space correlations.

\end{abstract}

\maketitle

{\emph {Introduction.---}} Electron interference turns momentum-space structure into observable
real-space patterns~\cite{gerry2005introductory}. Interference arises
from the superposition of different momentum components: their
separation determines the spatial oscillation period, while their
finite momentum spread determines the extent of the interference
pattern. In conventional solids, the relevant momentum and length
scales are usually set by band dispersion, as in the Fermi momentum
of Friedel oscillations~\cite{momentum} and the Fermi velocity entering
the superconducting coherence length
$\xi\sim\hbar v_F/\Delta$~\cite{vf}. Flat bands offer a different
limit. With dispersion largely quenched, spatial structure can instead
reflect the momentum dependence of the Bloch wavefunctions and,
ultimately, the geometry of the Hilbert space
~\cite{STIFF1,STIFF2,STIFF3,Pair4}.

The connection between quantum geometry and real-space structure is
well established through Wannier functions. The gauge-invariant spread
of maximally localized Wannier functions (MLWFs) is determined by the
Brillouin-zone (BZ) integral of the quantum metric
~\cite{MLWF1,MLWF2,MLWF3,MLWF4,MLWF5}. Related geometric length scales
also appear in superconducting correlations, Cooper-pair wavefunctions, and impurity bound states
~\cite{STIFF1,STIFF2,STIFF3,Pair1,Pair2,Pair3,Pair4,Pair5,jj3,jj}.
These results have motivated the quantum metric length as a measure of
the overall spatial extent
~\cite{qml2,qml3,qml4,qml5,Oscillation,QMLa,lower,lower2,qmp}.
However, integration over the BZ removes the momentum-resolved
information needed for interference. It does not tell us which momentum components form the long-distance tail, or how their characteristics determine its oscillation and decay.

In a flat band, the real-space structure follows from the Fourier
transform of the band projector. While its full momentum dependence
determines the short-range profile, the long-distance tail is dominated
by the nearest projector singularities in complex momentum space
~\cite{qml2,qml3}. Kohn showed that complex-momentum branch points
determine the exponential decay of Wannier functions~\cite{Kohn}, and
the same exponential decay associated with projector singularities has
recently been found for impurity bound states~\cite{qml2,embedding,cls,qml5}. A singularity
$k_\star=k_0+i\kappa$ contributes a long-distance component with
oscillation momentum $k_0$ and decay rate $\kappa$. Multiple
singularities therefore give multiple momentum components, whose
superposition produces real-space interference.

\begin{figure*}[t]
\centering
\includegraphics[width=0.9\linewidth]{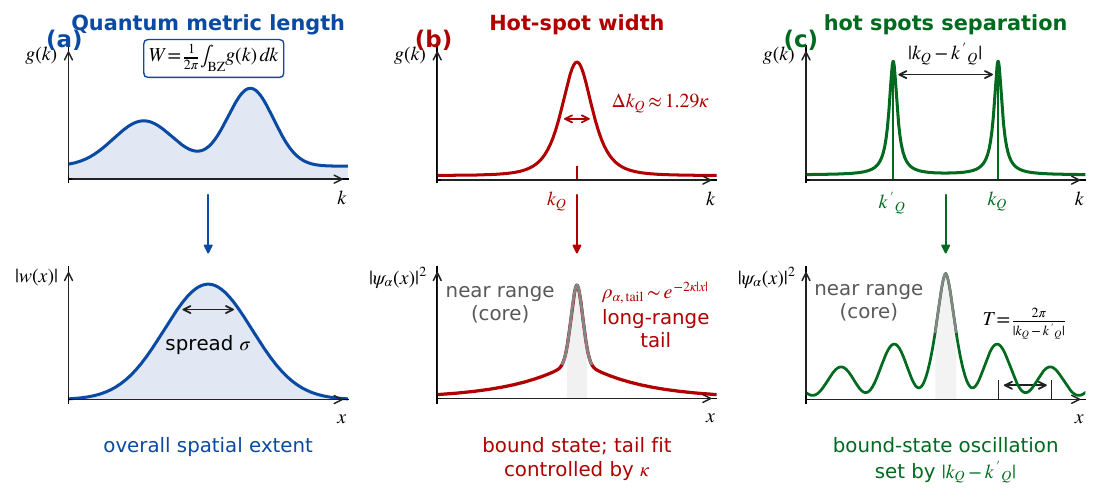}
\caption{
\textbf{Real-space information encoded in the quantum metric.}
(a) The quantum metric length sets the overall spread of the MLWF.
(b) The quantum-metric hot-spot width sets the asymptotic decay rate.
(c) The quantum-metric hot-spot separation sets the oscillation period.
}
\label{fig:hotspot_schematic}
\end{figure*}

Quantum-metric hot spots, where the metric is concentrated in momentum
space, have attracted attention for their roles in superfluid
stiffness~\cite{hu2025geometric,hot1}, charge-density oscillations, and bosonic-mode
squeezing~\cite{Oscillation,squeeze,Zeng2025}. Importantly, the momentum-resolved quantum metric is experimentally
accessible and has recently been mapped across the Brillouin zone
using Bloch-state tomography and ARPES~\cite{exp1,exp2},  providing experimental access to the position and width of quantum-metric hot spots. Here we show that these
hot spots also reveal the momentum components responsible for
real-space interference. As summarized in
Fig.~\ref{fig:hotspot_schematic}, the integrated quantum metric
characterizes the overall spatial extent
[Fig.~\ref{fig:hotspot_schematic}(a)], while individual hot spots
determine the long-distance interference pattern: their widths set the
decay length [Fig.~\ref{fig:hotspot_schematic}(b)], and their separation
sets the interference period [Fig.~\ref{fig:hotspot_schematic}(c)].
These relations follow from the nearest projector singularities, whose
imaginary and real parts determine the decay and oscillation scales,
respectively. In nearly flat bands, where dispersion-related scales are
suppressed, this geometric interference can also appear in coherent
transport, such as Josephson currents~\cite{jj1,jj2,jj4,jj5,jj6,jj7}.

{\emph {Projector singularities and quantum-metric.---}} We now relate the complex singularities of an isolated-band projector to
the long-distance behavior of bound states and the momentum-space
structure of the quantum metric. For a local weak impurity $\hat V (x)=\sum_\alpha {V}_\alpha \delta(x)$ satisfying $\operatorname{max}\{|{V}_\alpha|\} \ll \Delta$, where $\Delta$ is the band gap between the isolated band and adjacent bands, the bound-state wavefunctions projected onto the isolated band $\bar n$ reads~\cite{bound,jj3}
\begin{equation}
\psi_{\alpha}(x)
=
\frac{1}{N}
\sum_{k,\beta}
e^{ikx}
\frac{
P_{\bar{n}\alpha\beta}(k)
}{
E-\epsilon_{\bar n}(k)
}
V_{\beta}\psi_{\beta}(0),
\label{eq:projected_bound_state}
\end{equation}
where
$P_{\bar{n}\alpha\beta}(k)$ is an element in the band projector $P_{\bar{n}} = |u_{\bar n}(k)\rangle \langle u_{\bar n}(k)|$, and $\alpha,\beta$ label the orbitals.
For a dispersive band, the long-distance behavior of the bound state can be shaped by the complex singularities of both $P_{\bar n}(k)$ and
$[E-\epsilon_{\bar n}(k)]^{-1}$, which generally introduce different
oscillation and decay scales.

For an exactly flat band,
$\epsilon_{\bar n}(k)=\epsilon_{\mathrm{fb}}$, the energy denominator
is momentum independent, and the spatial dependence of the bound state
is determined solely by the Fourier transform of the band projector $P_{\bar n}(k)$.
Writing the normalized Bloch vector as
$|u_{\bar n}(k)\rangle=|v(k)\rangle/\sqrt{D(k)}$, where $|v(k)\rangle$ is the unnormalized part with no complex singularities and $D(k)=\langle v(k)|v(k)\rangle$, we can write the band projector in real space as
\begin{equation}
P_{\bar{n}\alpha\beta}(x)
=
\int_{\mathrm{BZ}}\frac{dk}{2\pi}\,
e^{ikx}
\frac{v_\alpha(k)v_\beta^{*}(k)}{D(k)} .
\label{eq:real_space_projector}
\end{equation}
The same real-space projector can be written in terms of Wannier
functions as
$P_{\bar{n}\alpha\beta}(x)
=
\sum_R
w_{\bar n\alpha}(x-R)
w_{\bar n\beta}^{*}(-R)$
\cite{lower,lower2}. For a non-topological band which is free from Wannier obstruction~\cite{brouder2007exponential}, the projected impurity states share the same spatial scales as the MLWFs of the band, while their orbital amplitudes and relative phases remain model dependent, as discussed in the Supplemental Material.

The quantum metric of the isolated band is
\begin{equation}
g_{\bar n}(k)
=
\langle\partial_k u_{\bar n}|\partial_k u_{\bar n}\rangle
-
|\langle u_{\bar n}|\partial_k u_{\bar n}\rangle|^2
=
\frac{f(k)}{D^2(k)},
\label{eq:quantum_metric}
\end{equation}
where the last form is derived in the Appendix. It is evident from Eqs.~\eqref{eq:real_space_projector} and \eqref{eq:quantum_metric} that $P_{\bar n}(k)$ and
$g_{\bar n}(k)$ have the same complex singularities associated with the zeros of
$D(k)$. These singularities determine the long-range behavior of
$P_{\bar n}(x)$ and, as shown below, the position and width of the
quantum-metric hot spots.

\begin{figure*}[t]
    \centering
    \includegraphics[width=1\textwidth]{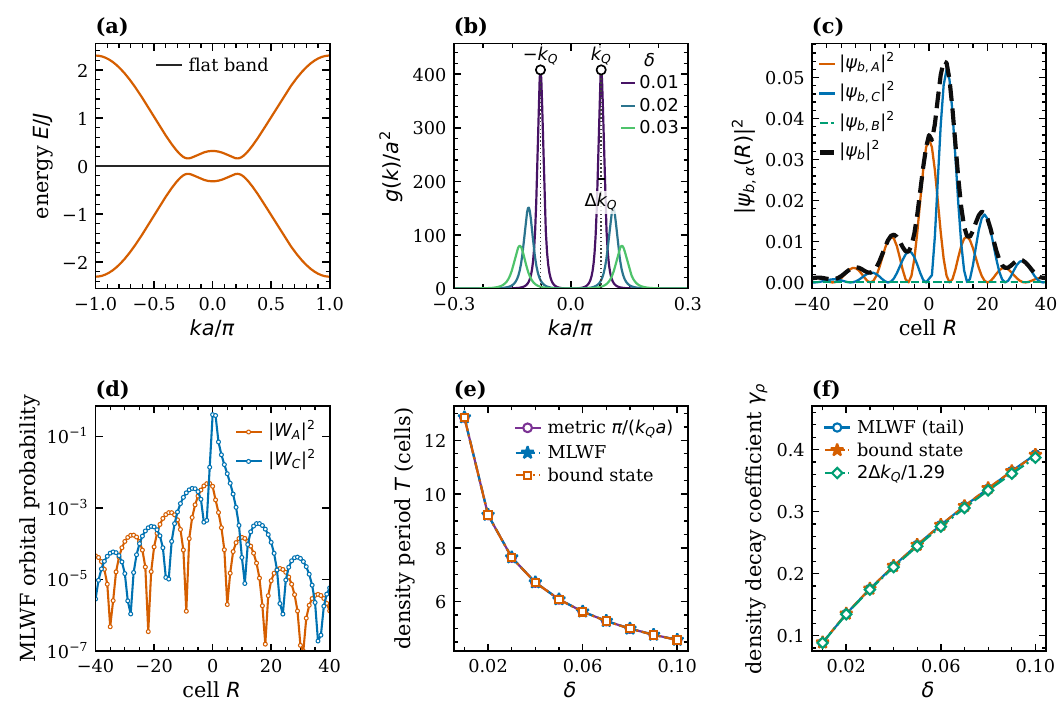}
    \caption{\textbf{Flat-band quantum geometry and real-space structure in the 1D Lieb lattice.}
    (a) Band structure at $\delta=0.1$, showing the flat band and the two
    dispersive bands.
    (b) Quantum metric $g_{\bar n}(k)$ for different $\delta$, exhibiting two
    symmetry-related hot spots at $\pm k_Q$ with FWHM $\Delta k_Q$.
    (c) Orbital-resolved probability distribution of the impurity bound state
    at $\delta=0.01$, showing the oscillatory long-range structure.
    (d) Corresponding MLWF probabilities on the $A$ and $C$ orbitals on a
    logarithmic scale, displaying the same oscillatory envelope.
    (e) Oscillation periods extracted from the MLWF and the $A$-orbital bound
    state, compared with the quantum-metric prediction $\pi/(k_Q a)$.
    (f) Long-range density decay coefficients of the MLWF and the $A$-orbital
    bound state, compared with $2\Delta k_Q/1.29$. Other parameters are: $J=1, \hat{V}(x)=V_0\delta(x) \operatorname{diag}([1,0,0])$ with $V_0=0.01\delta$ and $\delta'=5\delta$.}
    \label{fig:2}
\end{figure*}

Let $k_{\star,j}=k_{0,j}+i\kappa_j$ $(\kappa_j>0)$
denote the $j$-th singularity of the band projector in the complex
momentum plane. At long distances, the bound-state wavefunctions take
the asymptotic form
\begin{equation}
\psi_{\alpha}(x)
\sim
\sum_j
\mathcal A_{\alpha j}
e^{-\kappa_j|x|+ik_{0,j}x},
\label{eq:bound_state_asymptotic}
\end{equation}
As evident in Eq.~\eqref{eq:bound_state_asymptotic}, each singularity of the band projector defines an evanescent wave component in the bound-state wavefunctions, whose wavevector and decay length are set by the real and imaginary parts of the singularity, respectively. 
Taking the density distribution of a bound state, different evanescent components interfere with each other, producing a spatially oscillatory pattern, as illustrated in the lower panel of Fig.~\ref{fig:hotspot_schematic}(c).
Similarly, for multiple spatially overlapped bound states, the coupling strength between them also exhibits an oscillatory pattern, resulting in observable transport effects, as we will discuss later.

We refer to a pronounced and localized peak of the momentum-resolved
quantum metric as a \emph{quantum-metric hot spot}. Here, ``localized''
means that the peak forms a distinct feature on top of a background
that varies only weakly over its width, rather than being one of many
rapid oscillations of $g(k)$ across the Brillouin zone. Among such hot
spots, the correspondence discussed here applies when the peak is
well resolved and its local line shape is approximately
squared-Lorentzian. In this regime, the hot spot reflects the nearest
isolated projector singularity: its position is associated with the
real part of the singularity, while its width is associated with the
distance of the singularity from the real momentum axis.

For quantum-metric hot spots satisfying the above criteria, the local
metric profile is dominated by the nearest singularity of the band
projector. In this regime, the slowly varying regular prefactor can be
approximated by its value at $k_0$, giving
\begin{equation}
    g_{\bar n}(k)\simeq
    \frac{C(k_0)}
    {\left[(k-k_0)^2+\kappa^2\right]^2}.
\end{equation}
This gives
\begin{equation}
    k_Q \simeq k_0,
    \qquad
    \Delta k_Q
    \simeq
    2\sqrt{\sqrt{2}-1}\,\kappa
    \simeq 1.29\,\kappa ,
    \label{eq:metric_pole_relation}
\end{equation}
where $k_Q$ and $\Delta k_Q$ are the hot-spot position and full width at
half maximum (FWHM), respectively. Corrections due to the momentum
dependence of $C(k)$ and the nonsingular background are small when they
vary slowly over the hot-spot width, as shown in the Appendix

Here, $1.29$ denotes the numerical approximation to
$2\sqrt{\sqrt{2}-1}$ and is used in the figure labels for brevity.
We focus on the regime in which two well-separated hot spots, associated
with singularities $k_\star=k_0+i\kappa$ and
$k'_\star=k'_0+i\kappa$, dominate the long-distance behavior.
Their real-space contributions interfere with a relative wave vector
$|k_0-k'_0|$, giving an oscillation period
$2\pi/|k_0-k'_0|$. Since the hot-spot positions satisfy
$k_Q\simeq k_0$ and $k'_Q\simeq k'_0$, the hot-spot separation directly
encodes this interference wave vector, while the hot-spot width sets
the common decay rate $\kappa$. A narrower hot spot therefore
corresponds to a smaller decay rate and a more extended real-space tail.
The detailed derivation and the conditions for this correspondence are
given in the Appendix.
\begin{figure*}[t]
    \centering
    \includegraphics[width=1\linewidth]{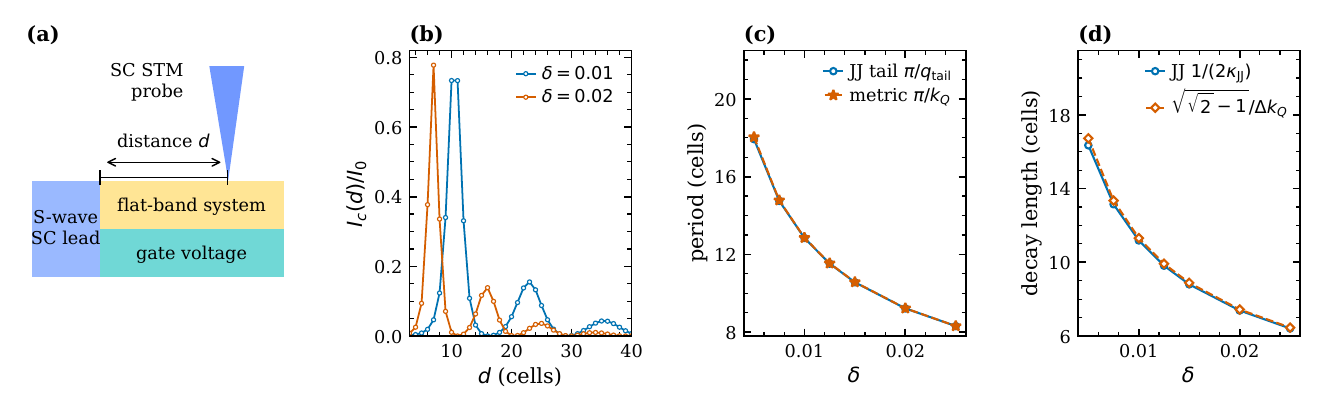}
   \caption{\textbf{Josephson transport in the 1D Lieb flat-band system at
    $k_BT/\Delta=0.01$.}
    (a) Junction geometry with two A-orbital contacts separated by $d$.
    (b) $I_c(d)\simeq|I(\pi/2,d)|$ for $\delta=0.01$ and $0.02$.
    (c) Long-range period $\pi/q_{\rm tail}$ compared with $\pi/k_Q$.
    (d) Decay length $1/(2\kappa_{\rm JJ})$ compared with
    $\sqrt{\sqrt{2}-1}/\Delta k_Q$.
   }
    \label{fig:5}
\end{figure*}

{\emph{Lieb-lattice model.---}}
We use a 1D Lieb lattice to test these relations. The conventional
Lieb lattice features a flat band with a single quantum metric hot spot~\cite{lieb1989two}, and the dominant projector singularities in the flat band lie near the high-symmetry points, i.e, $k=0$ or
$k=\pm\pi/a$. We adapt the Lieb lattice by introducing a hopping modulation between the A and B orbitals, which gives rise to multiple quantum metric hot spots in the preserved flat band. The projector singularities thus shift accordingly, as discussed in the previous section. The Hamiltonian is
\begin{equation}
H(k)
= J
\begin{pmatrix}
0 & \alpha(k) & 0 \\
\alpha^*(k) & 0 & \delta \\
0 & \delta & 0
\end{pmatrix},
\label{eq:lieb_model}
\end{equation}
with
\begin{equation}
\alpha(k)=1-\frac{1+\delta}{2}e^{ika}-\frac{1+\delta'}{2}e^{-ika}.
\end{equation}

Here, $a$ is the lattice constant, and the hopping modulation parameters $\delta, \delta'$ control the band structure as well as the quantum metric in the flat band. Figure~\ref{fig:2}(a) and (b) show the band structure and quantum metric
for different values of $\delta$. Two symmetry-related metric hot spots
at $\pm k_Q$ become sharper and more pronounced as $\delta$ decreases. Introducing a localized impurity $\hat{V}(x)$ at the zeroth site, one yields a localized bound state, which exhibits a spatially oscillatory feature, as shown in
Figs.~\ref{fig:2}(c) and (d). The impurity bound state exhibits pronounced
orbital-dependent oscillations [Fig.~\ref{fig:2}(c)], while the MLWF
displays corresponding oscillatory long-range tails
[Fig.~\ref{fig:2}(d)].

To quantify these real-space scales, Fig.~\ref{fig:2}(e) compares the
density oscillation periods extracted from the MLWF and the bound state
with the metric prediction $\pi/(k_Qa)$. The three agree over the full
parameter range. Likewise, Fig.~\ref{fig:2}(f) shows that their
long-range density decay coefficients are well captured by
$2\Delta k_Q/1.29$. Thus, the position and width of the metric hot spots
determine, respectively, the oscillation and decay scales of both the
MLWF and impurity-bound-state tails.

The same relation can be obtained analytically in a generalized
sawtooth model~\cite{SAW1,SAW2,SAW3,SAW4}, where the projector
singularities can be found explicitly; see the Supplemental Material.

{\emph{Application: Quantum metric and Josephson transport.---}}
Finally, we show that the real-space oscillations encoding the
momentum-resolved quantum metric can be probed through superconducting
transport. In an S/Lieb/S junction, each superconducting contact induces
a bound state in the flat-band region. The resulting Josephson oscillation
can be understood as arising from the overlap of these bound states, which
inherits their characteristic oscillation and decay~\cite{jj,jj3}. The distance dependence of the Josephson critical current
$I_c(d)$ can be measured using Josephson scanning tunneling microscopy
(STM)~\cite{STM1,STM2,STM3}, where a superconducting STM tip serves as
a moving contact, tuning the junction length $d$. The interference
between the bound states induced by the two contacts is thereby encoded
in $I_c(d)$ through its characteristic oscillation period and decay scale.

For the Josephson-junction geometry considered here, both superconducting
electrodes couple locally to the $A$ orbital of the flat-band region, with
the two contacts separated by a distance $d$. At low temperatures, the
projected propagator between the two contacts takes the asymptotic form
\begin{equation}
G_{AA}(d,i\omega)
\simeq
\mathcal Z(\omega)e^{-\kappa d}\cos\Theta_d,
\label{eq:josephson_projected_propagator}
\end{equation}
where
\begin{equation}
\Theta_d=k_0d-\theta_p,
\qquad
k_0=\operatorname{Re}k_\star,
\qquad
\kappa=\operatorname{Im}k_\star.
\end{equation}
Here, $k_\star$ is the nearest complex singularity of the flat-band
projector, $\theta_p$ is the phase shift set by the pole residue and orbital
structure, and $\mathcal Z(\omega)$ is a frequency-dependent prefactor
containing the residue and local orbital weights~\cite{jjs1,jjs2,jjs3}.
The effective coupling between the two contacts can be understood as the
overlap of the bound states induced from the two sides. Their oscillatory
tails interfere across the junction, giving an effective transmission
amplitude
\begin{equation}
t_{\rm eff}(d)
\simeq
t_0 e^{-\kappa d}\cos\Theta_d,
\label{eq:effective_transmission_main}
\end{equation}
where $t_0$ is a distance-independent prefactor. The corresponding
transmission probability is
\begin{equation}
\tau(d)
=
\frac{\tau_0}{2}e^{-2\kappa d}
\left[
1+\cos\left(2k_0d-2\theta_p\right)
\right],
\label{eq:tau_josephson_main}
\end{equation}
where $\tau_0$ sets the overall transmission scale. Thus, the interference
between the two bound states inherits the decay and oscillation set by the
same projector singularity.

In the low-temperature tunneling regime, $\tau(d)\ll1$, the critical current
is
\begin{equation}
I_c(d)
\simeq
\frac{e\Delta}{2\hbar}\tau(d),
\end{equation}
where $\Delta$ is the superconducting gap. Hence,
\begin{equation}
I_c(d)
\propto
e^{-2\kappa d}
\left[
1+\cos\left(2k_0d-2\theta_p\right)
\right].
\label{eq:critical_current_asymptotic}
\end{equation}

For generic orbital couplings, the phase and oscillation contrast depend
on the contacts, while the long-distance oscillation wave vector $2k_0$
and decay rate $2\kappa$ remain unchanged.

Figure~\ref{fig:5}(b) shows the decaying oscillations of the critical
current $I_c(d)$. As $\delta$ is varied, the fitted oscillation period
follows $\pi/k_Q$, while the decay length agrees with the scale obtained
from the width of the quantum-metric hot spot
[Fig.~\ref{fig:5}(c),(d)]. In the Lieb lattice, the zero of $f(k)$ at
$k=0$ lies close to, but outside, the hot-spot region, providing a
nontrivial test of the relation. 
 Detailed derivations of the effective transmission
are given in the Supplemental Material, together
with numerical results for a generalized sawtooth Josephson junction
that show the same oscillation and decay relations.

{\emph{Discussion and conclusion.---}}
In this work, we show that quantum-metric hot spots provide a
momentum-space picture of real-space interference in isolated nearly
flat bands. While the quantum metric length characterizes the overall
spread of the MLWF, individual hot spots are associated with the
long-distance momentum components of Wannier functions and bound
states. When isolated projector singularities dominate, each hot spot
tracks a corresponding complex-momentum component: its position sets
the oscillation momentum, while its width sets the exponential decay
rate. The superposition of components associated with different hot
spots then produces real-space interference, with the hot-spot
separation setting the interference period. The same interference and
decay scales appear in Wannier functions, impurity bound states, and
Josephson transport.

Our analysis applies to well-separated quantum-metric hot spots with a
local squared-Lorentzian profile, as expected when nearby projector
singularities dominate. When several hot spots are present, each is
associated with a long-distance momentum component. Their relative
positions determine the interference wave vectors, while their widths
set the decay scales of the corresponding components. More complex
real-space interference patterns can therefore arise from multiple hot
spots across the Brillouin zone.

The same picture can be extended to higher dimensions. In two
dimensions, one may fix one momentum component and Fourier transform
along the other direction. The resulting one-dimensional momentum
profile can then be analyzed in the same way as above: hot spots give
the dominant long-distance components, their separation sets the
interference period, and their widths set the decay scale along the
chosen direction. Repeating this analysis along different directions
can reveal the directional dependence of the real-space interference
pattern. A systematic study in higher dimensions is left for future
work.

\emph{Acknowledgements}---This work was supported by the Ministry of Science and Technology, China, The New Cornerstone Foundation, the State Key Laboratory of Quantum Information Technologies and Materials, and the Hong Kong Research Grants Council through
Grants No. MOST23SC01-A, No. RFS2021-6S03, No. 6C6053-23G, No. AoE/P-701/20, AoE/P-604/25R, No. 16309223, No. 16311424, and No. 16300325.

\nocite{*}
\bibliography{sorsamp}

@book{gerry2005introductory,
  title={Introductory quantum optics},
  author={Gerry, Christopher C and Knight, Peter L and Beck, Mark},
  volume={1},
  year={2005},
  publisher={Cambridge university press Cambridge}
}

@article{jj,
  title = {Superconducting junctions with flat bands},
  author = {Virtanen, P. and Penttil{\"a}, R. P. S. and T{\"o}rm{\"a}, P. and D{\'{\i}}ez-Carl{\'o}n, A. and Efetov, D. K. and Heikkil{\"a}, T. T.},
  journal = {Phys. Rev. B},
  volume = {112},
  issue = {10},
  pages = {L100502},
  numpages = {6},
  year = {2025},
  month = {Sep},
  publisher = {American Physical Society},
  doi = {10.1103/7f1y-kjyl},
  url = {https://link.aps.org/doi/10.1103/7f1y-kjyl}
}

@article{momentum,
     author = {Cristina Bena},
     title = {Friedel oscillations: {Decoding} the hidden physics},
     journal = {Comptes Rendus. Physique},
     pages = {302--321},
     year = {2016},
     publisher = {Elsevier},
     volume = {17},
     number = {3-4},
     doi = {10.1016/j.crhy.2015.11.006},
}

@book{vf,
  author = {Tinkham, Michael},
  month = jun,
  publisher = {Dover Publications},
  title = {Introduction to Superconductivity},
  url = {http://www.worldcat.org/isbn/0486435032},
  year = 2004
}

@article{MLWF1,
  title={Riemannian structure on manifolds of quantum states},
  author={J. P. Provost and G. Vall{\'e}e},
  journal={Communications in Mathematical Physics},
  year={1980},
  volume={76},
  pages={289-301},
  url={https://api.semanticscholar.org/CorpusID:120764117}
}

@article{MLWF2,
   title={Maximally localized generalized Wannier functions for composite energy bands},
   volume={56},
   ISSN={1095-3795},
   url={http://dx.doi.org/10.1103/PhysRevB.56.12847},
   DOI={10.1103/physrevb.56.12847},
   number={20},
   journal={Physical Review B},
   publisher={American Physical Society (APS)},
   author={Marzari, Nicola and Vanderbilt, David},
   year={1997},
   month=Nov, pages={12847–12865} }

@article{MLWF3,
   title={Maximally localized Wannier functions for entangled energy bands},
   volume={65},
   ISSN={1095-3795},
   url={http://dx.doi.org/10.1103/PhysRevB.65.035109},
   DOI={10.1103/physrevb.65.035109},
   number={3},
   journal={Physical Review B},
   publisher={American Physical Society (APS)},
   author={Souza, Ivo and Marzari, Nicola and Vanderbilt, David},
   year={2001},
   month=Dec }

@article{MLWF4,
   title={Maximally localized Wannier functions: Theory and applications},
   volume={84},
   ISSN={1539-0756},
   url={http://dx.doi.org/10.1103/RevModPhys.84.1419},
   DOI={10.1103/revmodphys.84.1419},
   number={4},
   journal={Reviews of Modern Physics},
   publisher={American Physical Society (APS)},
   author={Marzari, Nicola and Mostofi, Arash A. and Yates, Jonathan R. and Souza, Ivo and Vanderbilt, David},
   year={2012},
   month=Oct, pages={1419–1475} }

@article{MLWF5,
   title={The insulating state of matter: a geometrical theory},
   volume={79},
   ISSN={1434-6036},
   url={http://dx.doi.org/10.1140/epjb/e2010-10874-4},
   DOI={10.1140/epjb/e2010-10874-4},
   number={2},
   journal={The European Physical Journal B},
   publisher={Springer Science and Business Media LLC},
   author={Resta, R.},
   year={2011},
   month=Jan, pages={121–137} }

@article{STIFF1,
   title={Superfluidity in topologically nontrivial flat bands},
   volume={6},
   ISSN={2041-1723},
   url={http://dx.doi.org/10.1038/ncomms9944},
   DOI={10.1038/ncomms9944},
   number={1},
   journal={Nature Communications},
   publisher={Springer Science and Business Media LLC},
   author={Peotta, Sebastiano and Törmä, Päivi},
   year={2015},
   month=Nov }

@article{STIFF2,
   title={Geometric Origin of Superfluidity in the Lieb-Lattice Flat Band},
   volume={117},
   ISSN={1079-7114},
   url={http://dx.doi.org/10.1103/PhysRevLett.117.045303},
   DOI={10.1103/physrevlett.117.045303},
   number={4},
   journal={Physical Review Letters},
   publisher={American Physical Society (APS)},
   author={Julku, Aleksi and Peotta, Sebastiano and Vanhala, Tuomas I. and Kim, Dong-Hee and Törmä, Päivi},
   year={2016},
   month=July }

@article{STIFF3,
   title={Band geometry, Berry curvature, and superfluid weight},
   volume={95},
   ISSN={2469-9969},
   url={http://dx.doi.org/10.1103/PhysRevB.95.024515},
   DOI={10.1103/physrevb.95.024515},
   number={2},
   journal={Physical Review B},
   publisher={American Physical Society (APS)},
   author={Liang, Long and Vanhala, Tuomas I. and Peotta, Sebastiano and Siro, Topi and Harju, Ari and Törmä, Päivi},
   year={2017},
   month=Jan }

@article{Pair1,
   title={Anomalous coherence length in superconductors with quantum metric},
   volume={8},
   ISSN={2399-3650},
   url={http://dx.doi.org/10.1038/s42005-024-01930-0},
   DOI={10.1038/s42005-024-01930-0},
   number={1},
   journal={Communications Physics},
   publisher={Springer Science and Business Media LLC},
   author={Hu, Jin-Xin and Chen, Shuai A. and Law, K. T.},
   year={2025},
   month=Jan }

@article{Pair2,
   title={Correlation lengths of flat-band superconductivity from quantum geometry},
   volume={113},
   ISSN={2469-9969},
   url={http://dx.doi.org/10.1103/wfxg-5px1},
   DOI={10.1103/wfxg-5px1},
   number={21},
   journal={Physical Review B},
   publisher={American Physical Society (APS)},
   author={Elden, S. S. and Iskin, M.},
   year={2026},
   month=June }

@article{Pair3,
   title={Revisiting flat band superconductivity: Dependence on minimal quantum metric and band touchings},
   volume={106},
   ISSN={2469-9969},
   url={http://dx.doi.org/10.1103/PhysRevB.106.014518},
   DOI={10.1103/physrevb.106.014518},
   number={1},
   journal={Physical Review B},
   publisher={American Physical Society (APS)},
   author={Huhtinen, Kukka-Emilia and Herzog-Arbeitman, Jonah and Chew, Aaron and Bernevig, Bogdan A. and Törmä, Päivi},
   year={2022},
   month=July }

@article{Pair4,
  title = {Ginzburg-Landau Theory of Flat-Band Superconductors with Quantum Metric},
  author = {Chen, Shuai A. and Law, K. T.},
  journal = {Phys. Rev. Lett.},
  volume = {132},
  issue = {2},
  pages = {026002},
  numpages = {7},
  year = {2024},
  month = {Jan},
  publisher = {American Physical Society},
  doi = {10.1103/PhysRevLett.132.026002},
  url = {https://link.aps.org/doi/10.1103/PhysRevLett.132.026002}
}

@article{Pair5,
   title={Pair size and quantum geometry in a multiband Hubbard model},
   volume={111},
   ISSN={2469-9969},
   url={http://dx.doi.org/10.1103/PhysRevB.111.014502},
   DOI={10.1103/physrevb.111.014502},
   number={1},
   journal={Physical Review B},
   publisher={American Physical Society (APS)},
   author={Iskin, M.},
   year={2025},
   month=Jan }

@article{Weight1,
   title={Polarization and localization in insulators: Generating function approach},
   volume={62},
   ISSN={1095-3795},
   url={http://dx.doi.org/10.1103/PhysRevB.62.1666},
   DOI={10.1103/physrevb.62.1666},
   number={3},
   journal={Physical Review B},
   publisher={American Physical Society (APS)},
   author={Souza, Ivo and Wilkens, Tim and Martin, Richard M.},
   year={2000},
   month=July, pages={1666–1683} }

@article{Weight2,
   title={Electron Localization in the Insulating State},
   volume={82},
   ISSN={1079-7114},
   url={http://dx.doi.org/10.1103/PhysRevLett.82.370},
   DOI={10.1103/physrevlett.82.370},
   number={2},
   journal={Physical Review Letters},
   publisher={American Physical Society (APS)},
   author={Resta, Raffaele and Sorella, Sandro},
   year={1999},
   month=Jan, pages={370–373} }

@article{Weight4,
   title={Band geometry of fractional topological insulators},
   volume={90},
   ISSN={1550-235X},
   url={http://dx.doi.org/10.1103/PhysRevB.90.165139},
   DOI={10.1103/physrevb.90.165139},
   number={16},
   journal={Physical Review B},
   publisher={American Physical Society (APS)},
   author={Roy, Rahul},
   year={2014},
   month=Oct }

@article{Weight5,
   title={Relations between topology and the quantum metric for Chern insulators},
   volume={104},
   ISSN={2469-9969},
   url={http://dx.doi.org/10.1103/PhysRevB.104.045103},
   DOI={10.1103/physrevb.104.045103},
   number={4},
   journal={Physical Review B},
   publisher={American Physical Society (APS)},
   author={Ozawa, Tomoki and Mera, Bruno},
   year={2021},
   month=July }

@misc{Oscillation,
      title={Quantum Geometric Friedel Oscillations}, 
      author={Xing-Lei Ma and Jinchao Zhao and Bo-Qing Wu and Patrick A. Lee and K. T. Law},
      year={2026},
      eprint={2607.04654},
      archivePrefix={arXiv},
      primaryClass={cond-mat.mes-hall},
      url={https://arxiv.org/abs/2607.04654}, 
}

@article{jj1,
   title={Ballistic Josephson junctions in edge-contacted graphene},
   volume={10},
   ISSN={1748-3395},
   url={http://dx.doi.org/10.1038/nnano.2015.156},
   DOI={10.1038/nnano.2015.156},
   number={9},
   journal={Nature Nanotechnology},
   publisher={Springer Science and Business Media LLC},
   author={Calado, V. E. and Goswami, S. and Nanda, G. and Diez, M. and Akhmerov, A. R. and Watanabe, K. and Taniguchi, T. and Klapwijk, T. M. and Vandersypen, L. M. K.},
   year={2015},
   month=July, pages={761–764} }

@article{jj2,
   title={Quantum oscillations of the critical current and high-field superconducting proximity in ballistic graphene},
   volume={12},
   ISSN={1745-2481},
   url={http://dx.doi.org/10.1038/nphys3592},
   DOI={10.1038/nphys3592},
   number={4},
   journal={Nature Physics},
   publisher={Springer Science and Business Media LLC},
   author={Ben Shalom, M. and Zhu, M. J. and Fal’ko, V. I. and Mishchenko, A. and Kretinin, A. V. and Novoselov, K. S. and Woods, C. R. and Watanabe, K. and Taniguchi, T. and Geim, A. K. and Prance, J. R.},
   year={2015},
   month=Dec, pages={318–322} }

@article{SAW1,
   title={Ferromagnetism in the Hubbard model: Examples from models with degenerate single-electron ground states},
   volume={158},
   ISSN={1432-0916},
   url={http://dx.doi.org/10.1007/BF02108079},
   DOI={10.1007/bf02108079},
   number={2},
   journal={Communications in Mathematical Physics},
   publisher={Springer Science and Business Media LLC},
   author={Mielke, Andreas and Tasaki, Hal},
   year={1993},
   month=Nov, pages={341–371} }

@article{SAW2,
  title = {Ferromagnetism in the Hubbard models with degenerate single-electron ground states},
  author = {Tasaki, Hal},
  journal = {Phys. Rev. Lett.},
  volume = {69},
  issue = {10},
  pages = {1608--1611},
  numpages = {0},
  year = {1992},
  month = {Sep},
  publisher = {American Physical Society},
  doi = {10.1103/PhysRevLett.69.1608},
  url = {https://link.aps.org/doi/10.1103/PhysRevLett.69.1608}
}

@article{SAW3,
   title={Strongly correlated flat-band systems: The route from Heisenberg spins to Hubbard electrons},
   volume={29},
   ISSN={1793-6578},
   url={http://dx.doi.org/10.1142/S0217979215300078},
   DOI={10.1142/s0217979215300078},
   number={12},
   journal={International Journal of Modern Physics B},
   publisher={World Scientific Pub Co Pte Lt},
   author={Derzhko, Oleg and Richter, Johannes and Maksymenko, Mykola},
   year={2015},
   month=May, pages={1530007} }

@article{SAW4,
   title={Bose condensation in flat bands},
   volume={82},
   ISSN={1550-235X},
   url={http://dx.doi.org/10.1103/PhysRevB.82.184502},
   DOI={10.1103/physrevb.82.184502},
   number={18},
   journal={Physical Review B},
   publisher={American Physical Society (APS)},
   author={Huber, Sebastian D. and Altman, Ehud},
   year={2010},
   month=Nov }

@article{jj3,
  title = {Flat band Josephson junctions with quantum metric},
  author = {Li, Zhong C. F. and Deng, Yuxuan and Chen, Shuai A. and Efetov, Dmitri K. and Law, K. T.},
  journal = {Phys. Rev. Res.},
  volume = {7},
  issue = {2},
  pages = {023273},
  numpages = {7},
  year = {2025},
  month = {Jun},
  publisher = {American Physical Society},
  doi = {10.1103/PhysRevResearch.7.023273},
  url = {https://link.aps.org/doi/10.1103/PhysRevResearch.7.023273}
}

@article{qml2,
  title = {Majorana Zero Modes in the Lieb-Kitaev Model with Tunable Quantum Metric},
  author = {Guo, Xingyao and Ma, Xinglei and Ying, Xuzhe and Law, K. T.},
  journal = {Phys. Rev. Lett.},
  volume = {135},
  issue = {7},
  pages = {076601},
  numpages = {6},
  year = {2025},
  month = {Aug},
  publisher = {American Physical Society},
  doi = {10.1103/l3c7-knqm},
  url = {https://link.aps.org/doi/10.1103/l3c7-knqm}
}

@misc{qml3,
      title={Universal Boundary-Modes Localization from Quantum Metric Length}, 
      author={Xing-Lei Ma and Jin-Xin Hu and K. T. Law},
      year={2025},
      eprint={2509.05114},
      archivePrefix={arXiv},
      primaryClass={cond-mat.mes-hall},
      url={https://arxiv.org/abs/2509.05114}, 
}

@misc{qml4,
      title={Quantum Metric Length as a Fundamental Length Scale in Disordered Flat Band Materials}, 
      author={Chun Wang Chau and Tian Xiang and Shuai A. Chen and K. T. Law},
      year={2026},
      eprint={2602.01354},
      archivePrefix={arXiv},
      primaryClass={cond-mat.mes-hall},
      url={https://arxiv.org/abs/2602.01354}, 
}

@misc{qml5,
      title={Quantum Metric Bound State of Light}, 
      author={Jinchao Zhao and Rongning Liu and Xue-Yang Song and K. T. Law},
      year={2026},
      eprint={2606.22479},
      archivePrefix={arXiv},
      primaryClass={cond-mat.mes-hall},
      url={https://arxiv.org/abs/2606.22479}, 
}

@article{jj4,
   title={Short Ballistic Josephson Coupling in Planar Graphene Junctions with Inhomogeneous Carrier Doping},
   volume={120},
   ISSN={1079-7114},
   url={http://dx.doi.org/10.1103/PhysRevLett.120.077701},
   DOI={10.1103/physrevlett.120.077701},
   number={7},
   journal={Physical Review Letters},
   publisher={American Physical Society (APS)},
   author={Park, Jinho and Lee, Jae Hyeong and Lee, Gil-Ho and Takane, Yositake and Imura, Ken-Ichiro and Taniguchi, Takashi and Watanabe, Kenji and Lee, Hu-Jong},
   year={2018},
   month=Feb }

@article{jj5,
  title = {Universal limit of critical-current fluctuations in mesoscopic Josephson junctions},
  author = {Beenakker, C. W. J.},
  journal = {Phys. Rev. Lett.},
  volume = {67},
  issue = {27},
  pages = {3836--3839},
  numpages = {0},
  year = {1991},
  month = {Dec},
  publisher = {American Physical Society},
  doi = {10.1103/PhysRevLett.67.3836},
  url = {https://link.aps.org/doi/10.1103/PhysRevLett.67.3836}
}

@article{jj6,
   title={Superconducting quantum interference at the atomic scale},
   volume={18},
   ISSN={1745-2481},
   url={http://dx.doi.org/10.1038/s41567-022-01644-6},
   DOI={10.1038/s41567-022-01644-6},
   number={8},
   journal={Nature Physics},
   publisher={Springer Science and Business Media LLC},
   author={Karan, Sujoy and Huang, Haonan and Padurariu, Ciprian and Kubala, Björn and Theiler, Andreas and Black-Schaffer, Annica M. and Morrás, Gonzalo and Yeyati, Alfredo Levy and Cuevas, Juan Carlos and Ankerhold, Joachim and Kern, Klaus and Ast, Christian R.},
   year={2022},
   month=July, pages={893–898} }

@article{jj7,
   title={Single channel Josephson effect in a high transmission atomic contact},
   volume={3},
   ISSN={2399-3650},
   url={http://dx.doi.org/10.1038/s42005-020-00397-z},
   DOI={10.1038/s42005-020-00397-z},
   number={1},
   journal={Communications Physics},
   publisher={Springer Science and Business Media LLC},
   author={Senkpiel, Jacob and Dambach, Simon and Etzkorn, Markus and Drost, Robert and Padurariu, Ciprian and Kubala, Björn and Belzig, Wolfgang and Yeyati, Alfredo Levy and Cuevas, Juan Carlos and Ankerhold, Joachim and Ast, Christian R. and Kern, Klaus},
   year={2020},
   month=July }

@article{QMLa,
  title = {Identifying Instabilities with Quantum Geometry in Flat-Band Systems},
  author = {Zhang, Jia-Xin and Wang, Wen O. and Balents, Leon and Savary, Lucile},
  journal = {Phys. Rev. Lett.},
  volume = {136},
  issue = {17},
  pages = {176504},
  numpages = {8},
  year = {2026},
  month = {Apr},
  publisher = {American Physical Society},
  doi = {10.1103/rw6g-w7my},
  url = {https://link.aps.org/doi/10.1103/rw6g-w7my}
}

@article{embedding,
  title = {Embedding Independent Length Scale of Flat Bands},
  author = {Lee, Seokju and Lee, Seung Hun and Yang, Bohm-Jung},
  journal = {Phys. Rev. Lett.},
  volume = {137},
  issue = {1},
  pages = {016401},
  numpages = {6},
  year = {2026},
  month = {Jun},
  publisher = {American Physical Society},
  doi = {10.1103/tgtq-nrrt},
  url = {https://link.aps.org/doi/10.1103/tgtq-nrrt}
}

@article{Zeng2025,
   title={Quantum metric induced oscillations in nearly dispersionless flat bands},
   volume={111},
   ISSN={2469-9969},
   url={http://dx.doi.org/10.1103/PhysRevB.111.L121102},
   DOI={10.1103/physrevb.111.l121102},
   number={12},
   journal={Physical Review B},
   publisher={American Physical Society (APS)},
   author={Zeng, Hui and Zhou, Zijian and Duan, Wenhui and Huang, Huaqing},
   year={2025},
   month=Mar }

@article{Kohn,
    author = "Kohn, W.",
    title = "{Analytic Properties of Bloch Waves and Wannier Functions}",
    doi = "10.1103/PhysRev.115.809",
    journal = "Phys. Rev.",
    volume = "115",
    pages = "809--821",
    year = "1959"
}

@article{hot1,
  author  = {Liu, Le and Hong, Yu and Zhang, Cheng-Ping and Zhu, Jundong
             and Dong, Jingwei and Watanabe, Kenji and Taniguchi, Takashi
             and Du, Luojun and Shi, Dongxia and Law, K. T.
             and Zhang, Guangyu and Yang, Wei},
  title   = {Electric field tunable coupling strength and quantum metric hot spots
             in a moir{\'e} flatband superconductor},
  journal = {Nature Communications},
  volume  = {17},
  pages   = {8300},
  year    = {2026},
  doi     = {10.1038/s41467-026-74848-6}
}

@article{bound,
   title={Impurity-bound states and Green’s function zeros as local signatures of topology},
   volume={92},
   ISSN={1550-235X},
   url={http://dx.doi.org/10.1103/PhysRevB.92.085126},
   DOI={10.1103/physrevb.92.085126},
   number={8},
   journal={Physical Review B},
   publisher={American Physical Society (APS)},
   author={Slager, Robert-Jan and Rademaker, Louk and Zaanen, Jan and Balents, Leon},
   year={2015},
   month=Aug }

@article{STM1,
   title={Scanning Josephson spectroscopy on the atomic scale},
   volume={93},
   ISSN={2469-9969},
   url={http://dx.doi.org/10.1103/PhysRevB.93.161115},
   DOI={10.1103/physrevb.93.161115},
   number={16},
   journal={Physical Review B},
   publisher={American Physical Society (APS)},
   author={Randeria, Mallika T. and Feldman, Benjamin E. and Drozdov, Ilya K. and Yazdani, Ali},
   year={2016},
   month=Apr }

@article{STM2,
   title={Josephson scanning tunneling microscopy},
   volume={64},
   ISSN={1095-3795},
   url={http://dx.doi.org/10.1103/PhysRevB.64.212506},
   DOI={10.1103/physrevb.64.212506},
   number={21},
   journal={Physical Review B},
   publisher={American Physical Society (APS)},
   author={Šmakov, Jurij and Martin, Ivar and Balatsky, Alexander V.},
   year={2001},
   month=Nov }

@article{STM3,
   title={Josephson scanning tunneling microscopy: A local and direct probe of the superconducting order parameter},
   volume={80},
   ISSN={1550-235X},
   url={http://dx.doi.org/10.1103/PhysRevB.80.144506},
   DOI={10.1103/physrevb.80.144506},
   number={14},
   journal={Physical Review B},
   publisher={American Physical Society (APS)},
   author={Kimura, Hikari and Barber, R. P. and Ono, S. and Ando, Yoichi and Dynes, R. C.},
   year={2009},
   month=Oct }

@article{jjs1,
author = {Caroli, C and Combescot, R and Nozieres, P and Saint-James, D},
year = {1971},
month = {03},
pages = {916-929},
title = {Direct calculation of the tunneling current},
volume = {4},
journal = {Journal of Physics C: Solid State Physics},
doi = {10.1088/0022-3719/4/8/018}
}

@article{jjs2,
  title = {Relation between conductivity and transmission matrix},
  author = {Fisher, Daniel S. and Lee, Patrick A.},
  journal = {Phys. Rev. B},
  volume = {23},
  issue = {12},
  pages = {6851(R)--6854(R)},
  numpages = {0},
  year = {1981},
  month = {Jun},
  publisher = {American Physical Society},
  doi = {10.1103/PhysRevB.23.6851},
  url = {https://link.aps.org/doi/10.1103/PhysRevB.23.6851}
}

@article{jjs3,
  title = {Universal limit of critical-current fluctuations in mesoscopic Josephson junctions},
  author = {Beenakker, C. W. J.},
  journal = {Phys. Rev. Lett.},
  volume = {67},
  issue = {27},
  pages = {3836--3839},
  numpages = {0},
  year = {1991},
  month = {Dec},
  publisher = {American Physical Society},
  doi = {10.1103/PhysRevLett.67.3836},
  url = {https://link.aps.org/doi/10.1103/PhysRevLett.67.3836}
}

@misc{lower,
      title={Quantum Geometric Kondo Cloud}, 
      author={Grant Z. X. Yang and K. T. Law},
      year={2026},
      eprint={2608.14072},
      archivePrefix={arXiv},
      primaryClass={cond-mat.str-el},
      url={https://arxiv.org/abs/2608.14072}, 
}

@misc{lower2,
      title={Impurity-dependent quantum geometry in 2D flat band superconductors}, 
      author={Simão S. Cardoso and A. Mesaros and P. Simon},
      year={2026},
      eprint={2609.00338},
      archivePrefix={arXiv},
      primaryClass={cond-mat.supr-con},
      url={https://arxiv.org/abs/2609.00338}, 
}

@article{cls,
  title = {Real-space decay of flat-band projectors from compact localized states},
  author = {Kim, Yeongjun and Flach, Sergej and Andreanov, Alexei},
  journal = {Phys. Rev. B},
  volume = {113},
  issue = {23},
  pages = {235110},
  numpages = {12},
  year = {2026},
  month = {Jun},
  publisher = {American Physical Society},
  doi = {10.1103/x5rg-lfkb},
  url = {https://link.aps.org/doi/10.1103/x5rg-lfkb}
}

@misc{qmp,
      title={Quantum Metric Localization and Quantum Metric Protection}, 
      author={Wen-Bo Dai and Jinchao Zhao and Shuai A. Chen and K. T. Law},
      year={2026},
      eprint={2605.03987},
      archivePrefix={arXiv},
      primaryClass={cond-mat.mes-hall},
      url={https://arxiv.org/abs/2605.03987}, 
}

@misc{squeeze,
      title={Quantum-Geometric Length Scale for Long-distance Squeezing in Bosonic Bogoliubov Systems}, 
      author={Sonu Verma and Sangmo Cheon and Myung-Joong Hwang and Moon Jip Park},
      year={2026},
      eprint={2608.23380},
      archivePrefix={arXiv},
      primaryClass={quant-ph},
      url={https://arxiv.org/abs/2608.23380}, 
}

@article{lieb1989two,
  title={Two theorems on the Hubbard model},
  author={Lieb, Elliott H},
  journal={Physical review letters},
  volume={62},
  number={10},
  pages={1201},
  year={1989},
  publisher={APS}
}

@article{hu2025geometric,
  title={Geometric and conventional contributions of superconducting diode effect: Application to flat-band systems},
  author={Hu, Jin-Xin and Chen, Shuai A and Law, KT},
  journal={Physical Review B},
  volume={111},
  number={17},
  pages={174513},
  year={2025},
  publisher={APS}
}

@article{exp1,
  title = {Extracting the quantum geometric tensor of an optical Raman lattice by Bloch-state tomography},
  author = {Yi, Chang-Rui and Yu, Jinlong and Yuan, Huan and Jiao, Rui-Heng and Yang, Yu-Meng and Jiang, Xiao and Zhang, Jin-Yi and Chen, Shuai and Pan, Jian-Wei},
  journal = {Phys. Rev. Res.},
  volume = {5},
  issue = {3},
  pages = {L032016},
  numpages = {6},
  year = {2023},
  month = {Aug},
  publisher = {American Physical Society},
  doi = {10.1103/PhysRevResearch.5.L032016},
  url = {https://link.aps.org/doi/10.1103/PhysRevResearch.5.L032016}
}

@article{
exp2,
author = {Sunje Kim  and Yoonah Chung  and Yuting Qian  and Soobin Park  and Chris Jozwiak  and Eli Rotenberg  and Aaron Bostwick  and Keun Su Kim  and Bohm-Jung Yang },
title = {Direct measurement of the quantum metric tensor in solids},
journal = {Science},
volume = {388},
number = {6751},
pages = {1050-1054},
year = {2025},
doi = {10.1126/science.ado6049},
URL = {https://www.science.org/doi/abs/10.1126/science.ado6049},
eprint = {https://www.science.org/doi/pdf/10.1126/science.ado6049},
}

@article{brouder2007exponential,
  title={Exponential localization of Wannier functions in insulators},
  author={Brouder, Christian and Panati, Gianluca and Calandra, Matteo and Mourougane, Christophe and Marzari, Nicola},
  journal={Physical review letters},
  volume={98},
  number={4},
  pages={046402},
  year={2007},
  publisher={APS}
}

\section*{Appendix}
\subsection{Quantum metric and projector singularities}

Here we derive the relation between a nearby complex projector
singularity and the position and width of a quantum-metric hot spot.

We write the normalized Bloch state as
\begin{equation}
|u(k)\rangle
=
\frac{|v(k)\rangle}{\sqrt{D(k)}},
\qquad
|v(k)\rangle
=
\begin{pmatrix}
A_1(k)\\
A_2(k)\\
\vdots\\
A_{N_{\rm orb}}(k)
\end{pmatrix},
\label{eq:unnormalized_spinor}
\end{equation}
where
\begin{equation}
D(k)
=
\langle v(k)|v(k)\rangle
=
\sum_{\alpha}
A_{\alpha}^{*}(k)A_{\alpha}(k).
\label{eq:D_definition}
\end{equation}
For an isolated band, $D(k)>0$ on the real Brillouin zone. The
one-dimensional quantum metric is
\begin{equation}
g(k)
=
\langle\partial_k u|\partial_k u\rangle
-
\left|
\langle u|\partial_k u\rangle
\right|^2
=
\frac{f(k)}{D^2(k)},
\label{eq:metric_unnormalized}
\end{equation}
with
\begin{equation}
f(k)
=
D(k)\sum_{\alpha}
|\partial_k A_{\alpha}(k)|^2
-
\left|
\sum_{\alpha}
A_{\alpha}^{*}(k)\partial_k A_{\alpha}(k)
\right|^2
\geq 0 .
\label{eq:Q_definition}
\end{equation}

For a short-range tight-binding model, the components $A_\alpha(k)$
are finite or rapidly convergent Fourier series,
\begin{equation}
A_{\alpha}(k)
=
a_{\alpha}
+
\sum_{m\geq1}
\left(
b_{\alpha,m}^{+}e^{imka}
+
b_{\alpha,m}^{-}e^{-imka}
\right).
\label{eq:An}
\end{equation}
Their analytic continuations are therefore regular in a finite region
of complex momentum. The singularities of the normalized projector
arise from complex zeros of the normalization factor $D(k)$, provided
that they are not canceled by the regular numerator.

Consider the nearest conjugate pair of zeros,
\begin{equation}
k_{\star}=k_0+i\kappa,
\qquad
k_{\star}^{*}=k_0-i\kappa,
\qquad
\kappa>0.
\end{equation}
Near this pair, $D(k)$ can be factorized as
\begin{equation}
D(k)
=
d(k)
(k-k_{\star})(k-k_{\star}^{*})
\label{eq:D_factorization}
\end{equation}
where $d(k)$ is regular and nonzero near $k_0$. The quantum metric can
then be written as
\begin{equation}
g(k)
\simeq
\frac{C(k)}
{\left[(k-k_0)^2+\kappa^2\right]^2},
\qquad
C(k)=\frac{f(k)}{d^2(k)} ,
\label{eq:metric_local}
\end{equation}
where $C(k)$ is regular near the hot spot. If $C(k)$ varies weakly over
the scale $\kappa$, it can be replaced by $C(k_0)$, giving
\begin{equation}
g_0(k)
=
\frac{C(k_0)}
{\left[(k-k_0)^2+\kappa^2\right]^2}.
\label{lorentz}
\end{equation}
The peak is therefore centered at
\begin{equation}
k_Q=k_0,
\end{equation}
and its full width at half maximum is
\begin{equation}
\Delta k_Q
=
2\sqrt{\sqrt{2}-1}\,\kappa
\simeq 1.29\,\kappa .
\label{eq:ideal_peak}
\end{equation}
\subsection{Corrections to the Hot-Spot Position and Width}

We consider the momentum dependence of the regular prefactor
$C(k)$ in Eq.~\eqref{eq:metric_local}. Defining
\begin{equation}
F(k)=\ln C(k),
\qquad
x=\frac{k-k_0}{\kappa},
\end{equation}
we expand around $k_0$ as
\begin{equation}
F(k_0+\kappa x)
=
F(k_0)
+
Ax
+
\frac{B}{2}x^2
+\cdots ,
\label{eq:F_expansion}
\end{equation}
where
\begin{equation}
A
=
\kappa
\left.
\frac{C'(k)}{C(k)}
\right|_{k_0},
\qquad
B
=
\kappa^2
\left.
\frac{d^2\ln C(k)}{dk^2}
\right|_{k_0}.
\label{eq:AB_definition}
\end{equation}
The dimensionless parameters $A$ and $B$ measure the local variation
of the regular prefactor over the momentum scale $\kappa$. We first
assume $|A|,|B|\ll1$ and discuss this condition for short-range
tight-binding models below. The local metric profile then becomes
\begin{equation}
\ln g(x)
\simeq
F(k_0)
+
Ax
+
\frac{B}{2}x^2
-
2\ln(1+x^2).
\label{eq:logg_local}
\end{equation}

The peak position,
\begin{equation}
x_Q=\frac{k_Q-k_0}{\kappa},
\end{equation}
is determined by $\partial_x\ln g(x_Q)=0$, which gives
\begin{equation}
A
=
x_Q
\left[
\frac{4}{1+x_Q^2}
-
B
\right].
\label{eq:peak_implicit}
\end{equation}
For small $A$ and $B$, the root connected continuously to the
unperturbed peak at $x_Q=0$ is
\begin{equation}
\frac{k_Q-k_0}{\kappa}
=
x_Q
\simeq
\frac{A}{4-B}.
\label{eq:peak_shift}
\end{equation}

We now consider the FWHM. The half-maximum points $x_\pm$ satisfy
\begin{equation}
\ln g(x_\pm)
=
\ln g(x_Q)-\ln2.
\label{eq:halfmax_condition}
\end{equation}
For $A=B=0$, they are $x_\pm=\pm c$. Writing
$x_\pm=\pm c+\delta_\pm$ and expanding about these points gives, to
second order in $A$ and first order in $B$,
\begin{equation}
x_\pm
\simeq
\pm c
+
\frac{\sqrt{2}}{4}A
\pm
\frac{c}{4\sqrt{2}}
\left(
B+\frac{3}{4}A^2
\right).
\label{eq:xpm_final}
\end{equation}
The term linear in $A$ shifts both half-maximum points equally and
therefore cancels from their separation. Defining the unperturbed FWHM
as
\begin{equation}
\Delta k_0
=
2c\kappa
=
2\sqrt{\sqrt{2}-1}\,\kappa,
\end{equation}
the relative correction to the width is
\begin{equation}
\frac{\Delta k_Q-\Delta k_0}{\Delta k_0}
\simeq
\frac{\sqrt{2}}{8}
\left(
B+\frac{3}{4}A^2
\right).
\label{eq:width_correction}
\end{equation}
\begin{figure}
    \centering
    \includegraphics[width=0.9\linewidth]{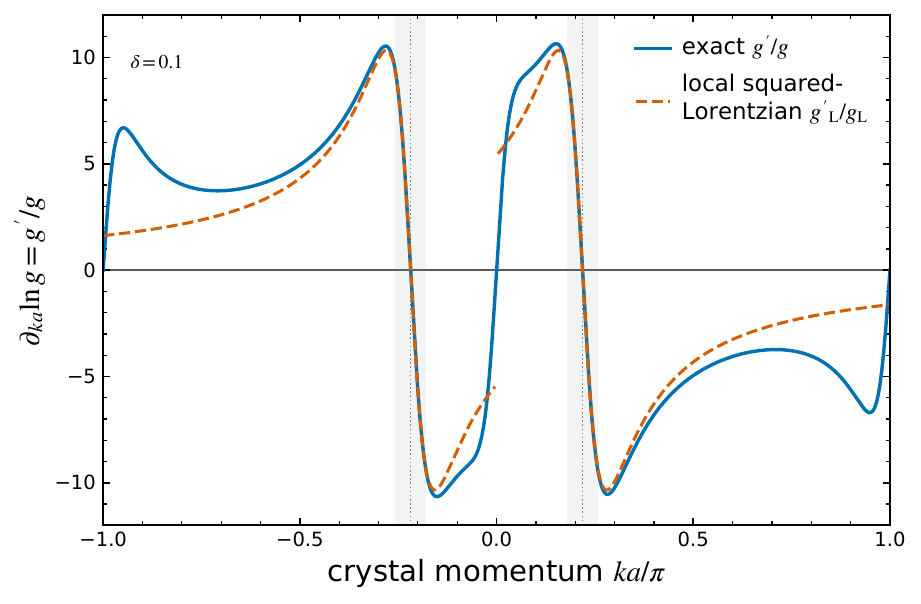}
    \caption{
    Comparison between $g'/g$ and the squared-Lorentzian result $g'_0/g_0$
    for the Lieb lattice at $\delta=0.1$. The shaded regions indicate the
    ranges within the full width at half maximum (FWHM) of  the quantum-metric
    hot spots. The divergences at $k=0,\pm\pi$ arise from the zeros of $g(k)$.
    }
    \label{fig:4}
\end{figure}
\subsection{Slow Variation of the Regular Prefactor}

For short-range tight-binding models, the functions entering $f(k)$
are finite or rapidly convergent Fourier series. For simple few-orbital models, the low-order harmonics usually dominate,
so that $f(k)$ varies on the Brillouin-zone scale. The regular factor
$d(k)$ in Eq.~(\ref{eq:D_factorization}) is also smooth near $k_0$ when
the nearest singularity is well separated from the remaining
singularities. The combined prefactor
\begin{equation}
C(k)=\frac{f(k)}{d^2(k)}
\end{equation}
therefore typically varies on a momentum scale of order $\pi/a$, when $f(k)$ does not become anomalously small near the hot spot,
\begin{equation}
A=O(\kappa a),
\qquad
B=O[(\kappa a)^2],
\end{equation}
and a narrow hot spot with $\kappa a\ll1$ naturally gives
$|A|,|B|\ll1$.

When $C(k)$ varies weakly over the scale $\kappa$, a nearby quadratic
projector singularity gives the squared-Lorentzian form of
Eq.~(\ref{lorentz}); variations on the same scale instead shift and
distort the peak [Eqs.~(\ref{eq:peak_shift}) and
(\ref{eq:width_correction})]. This can be checked directly by writing
$g(k)=C(k)g_0(k)$, such that
$g'/g=g'_0/g_0+C'/C$. As shown in Fig.~\ref{fig:4}, the close agreement
between $g'/g$ and $g'_0/g_0$ near the hot spots confirms that the
regular contribution $C'/C$ is small.

\end{document}